\documentclass[a4paper]{article}

\usepackage{INTERSPEECH2018}
\usepackage{algorithmic}
\usepackage{algorithm}
\usepackage{array}
\usepackage{multirow}
\newcommand\Tstrut{\rule{0pt}{2.6ex}}
 \usepackage{subcaption}
\usepackage{xcolor}
\title{Towards Automated Single Channel Source Separation using Neural Networks}
\name{Arpita Gang$^1$, Pravesh Biyani $^1$, Akshay Soni $^2$}
\address{
  $^1$IIIT Delhi, India\\
  $^2$Oath Research, USA}
\email{\{arpita1478, praveshb\}@iiitd.ac.in, akshaysoni@oath.com}

\begin{document}
\maketitle
\begin{abstract}
  Many applications of single channel source separation (SCSS) including automatic speech recognition (ASR), hearing aids etc.  require an estimation of only one source from a mixture of many sources.  Treating this special case as a regular SCSS problem where in all constituent sources are given equal priority in terms of reconstruction may result in a suboptimal separation performance. In this paper, we tackle the one source separation problem by suitably modifying the orthodox SCSS framework and focus only on one source at a time.  The proposed approach is a generic framework that can be applied to any existing SCSS algorithm, improves performance, and scales well when there are more than two sources in the mixture unlike most existing SCSS methods. Additionally, existing SCSS algorithms rely on fine hyper-parameter tuning hence making them difficult to use in practice. Our framework takes a step towards automatic tuning of the hyper-parameters thereby making our method better suited for the mixture to be separated and thus practically more useful. We test our framework on a neural network based algorithm and the results show an improved performance in terms of SDR and SAR.
\end{abstract}

\section{Introduction}
Single Channel Source Separation (SCSS) is an extraction of two or more underlying source components from a single observation of their mixture.  The SCSS problem arises in many scenarios including speech-denoising in telephony (e.g in call centers), for automatic speech recognition (ASR), sound separation as a preprocessing for hearing aids specially in non-stationary environments, and even sound event detection like fire-alarm, scream detection in security related applications. \par 
Recent research works \cite{nmf_deconv, nmf_sparse, adap_sparsity, dnn1, joint_mask} on supervised SCSS problem have improved the more traditional techniques of unsupervised blind source separation (BSS) by incorporating the source training data to generate models, both linear and non-linear, resulting in effective separation of sources.  In particular, models based on Deep Neural Networks (DNN) have had a tremendous success in the source separation tasks and pushed the performance to a place where now source separation may act as a practically useful pre-processing task. \par 
Most of the above mentioned applications require separation of one dominant source from the mixture containing audio signal from multiple sources. For instance, in the hearing aid scenario, generally one speech signal needs to be separated from the mixture containing one or more `noise' sources. Similar observation holds when source separation is applied as a preprocessing task in ASR. \par
While many SCSS setups require one main source to be separated, most existing model based source separation methods aim at simultaneous extraction of all the sources. Here each individual source model is required to not only mitigate interference by other sources but also  provide reasonable reconstruction performance. Effectively, a single optimization formulation when burdened with providing equal priority to all sources, results in a suboptimal performance for every source. This issue becomes more grave as the number of sources increase. \par
\subsection{Contributions}
Motivated by relevant applications and issues with joint separation paradigm, we introduced a `one source at a time' separation framework in \cite{dfnmf} for two sources when non-negative matrix factorization (NMF) based models for each source is used.   The first contribution of this paper is the extension of this strategy to any number of sources when the sources are modeled using DNN. We combine all the sources other than the concerned one into a single source, which we term as 'interferer'.  This converts the multi-source separation problem into a two-source separation problem where we concentrate on effective separation of only one source at the cost of other interfering sources. One key step is the inclusion of a term in the objective function that increases the distance between the source estimate and interferer signal in its orthogonal component thereby effectively increasing the source energy to artifact energy ratio (SAR) while maintaining the a good source to interference ratio (SIR).  Finally, our approach, which we call Discriminative Framework for DNN (DF-DNN), is generic enough and can be applied to many source separation setups. \par
Breaking a multiple source separation problem into two source separation problem for each source also assists in hyperparameter tuning, especially when DNNs are used. Generally, hyper parameters are tuned by employing a hit and trial on many parameter values using the development data set, at times even using manual intervention. In this paper, we attempt to make the parameter tuning systematic by defining few meta-parameters that act as a proxy for the key separation performance indices. By observing these parameters during the training, we can arrive at the appropriate parameter values and achieve training models that have a higher likelihood of providing good separation performance. Note that our meta-parameters--like our framework--is generic in nature and can be utilised over many discriminative source separation frameworks.
\subsection{Related Work}
Most recent works in SCSS have employed methods that learn discriminative models that utilize the training data to represent individual sources such that a source model represents itself well while simultaneously acts as a poor fit for other sources. \cite{joint_mask, disc_enhancement}.  The work in \cite{nmf_cc} proposed an regularized formulation that jointly trains the NMF dictionaries penalizing the coherence between trained models. Methods proposed in \cite{nmf_disc} and \cite{disc_rnn} aim at directly optimizing the SNR while training NMF dictionaries and recurrent neural networks respectively. The work proposed in \cite{joint_mask} trains a deep recurrent neural network that optimizes the estimated masks of the sources, while \cite{disc_enhancement} discriminatively enhances the separated sources. To the best of our knowledge researchers have not focused on automated hyper-parameter tuning in their works.
\section{Problem Description}
Single channel source separation requires the estimation of $L$ sources from a single observation of their mixture
\begin{equation}\label{scss}
x(t) = \sum_{i=1}^{L}y_i(t),
\end{equation}
where $y_{i}(t),~i = 1 \dots L$ is the $i^{th}$ source to be estimated and $x(t)$ is the observed mixture. Assuming that sources belong to subspaces in the ambient vector space, the difficulty of separation increases when the sources share basis of subspaces they belong to. On the other hand, when the sources are represented by orthogonal subspaces, increasing $L$ does not have any impact, while if the subspaces have large overlap, the separation performance decreases with an increase in number of sources. This is a limitation of traditional source separation methods like \cite{joint_mask,disc_enhancement, nmf_cc, disc_rnn} that focus on simultaneous reconstruction of all the $L$ sources. \par
On the other hand, the proposed \emph{one source at a time} approach decomposes the $L$-source separation problem into $L$ different source separation problems each focusing on the estimation of just one source and treating all the other sources as interference. Concretely, for the estimation of first source, i.e., $y_{1}(t)$ we consider all the other sources, $y_i(t),~i=2 \dots L$, as interference. Effectively, we have $L$ source separation problems, one for each source. This leads to high quality estimation of each source as empirically demonstrated later. 
Without loss of generality, we will look at the problem of estimating the first source from the mixture. To that end, we unfold \eqref{scss} as
\begin{equation}\label{scss1}
x(t) =\underbrace{y_1(t)}_\text{source} + \underbrace{(y_2(t) + ... + y_L(t))}_\text{interferers},
\end{equation}
where above we emphasize that all the sources other than $y_1(t)$ combined together act as interferer effectively reducing the multiple source separation problem into a two source problem. 
The aim of source separation is to estimate the underlying sources in a way that each source is reconstructed with little deformity and has minimal traces of other sources. The errors in an estimated source is defined are \cite{perf_eval}. Let $\mathbb{P}[y_1, y_2, \dots, y_k]$ be the orthogonal projector onto the space spanned by the vectors $y_1, y_2, \dots, y_k$, then the interference and artifacts introduced in the estimate of $j^{\rm{th}}$ source, denoted by $\hat{y}_j$, are given by 
\begin{eqnarray} \label{Interferer_Power}
E_{I}(j) &=& [\mathbb{P}_{I} - \mathbb{P}_{S}(j)]\hat{y}_j \\
E_{A}(j) &=& \hat{y}_j - \mathbb{P}_{I}\hat{y}_j
\end{eqnarray}
where 
\begin{equation*}
\mathbb{P}_{I} = \mathbb{P}[(y_{j'})_{1 \leq j' \leq L}]
\end{equation*}
and 
\begin{equation*}
\mathbb{P}_S(j) = \mathbb{P}[y_{j}].
\end{equation*}
The SIR and SAR of estimated source $\hat{y}_j$ are defined by
\begin{eqnarray}\label{sir}
\text{SIR} = 10\log_{10}\frac{\|\mathbb{P}_S(j)\hat{y}_j \|^2}{\|E_{I}(j)\|^2}\\
\text{SAR} = 10\log_{10}\frac{\|\mathbb{P}_I\hat{y}_j \|^2}{\|E_{A}(j)\|^2} \label{sar}
\end{eqnarray}
% \textcolor{red}{This lines needs to be expanded better.} 
It is evident from \eqref{Interferer_Power} that the interference $E_{I}$ is the component of the source estimate that lies in the space orthogonal to $\mathbb{P}_S$. The lesser the energy in $E_{I}$, better is the SIR. On the other hand we are not concerned to remove the part of interferer that lies in the subspace $\mathbb{P}_S(j)$ of the source while separating it. Therefore our strategy gives us the freedom to focus on the reconstruction of the source at the cost of interferer by subsuming overlapping subspace of interferer and source to $\mathbb{P}_S(j)$.  This means that the source can use the overlapping part of the interferer for its reconstruction that can increase the SAR \eqref{sar} of the estimated source.

The above insight is incorporated in our objective function (in Section \ref{sec:algo}) by adding a corresponding fit term for the component of interferer orthogonal to source. The generality of our approach makes it applicable to all kinds of supervised discriminative methods. In this paper, we apply our framework on a neural network based model described in \cite{joint_mask}. 
\begin{figure}
	 \centering
	\includegraphics[height=7.5cm,width=1.1\linewidth]{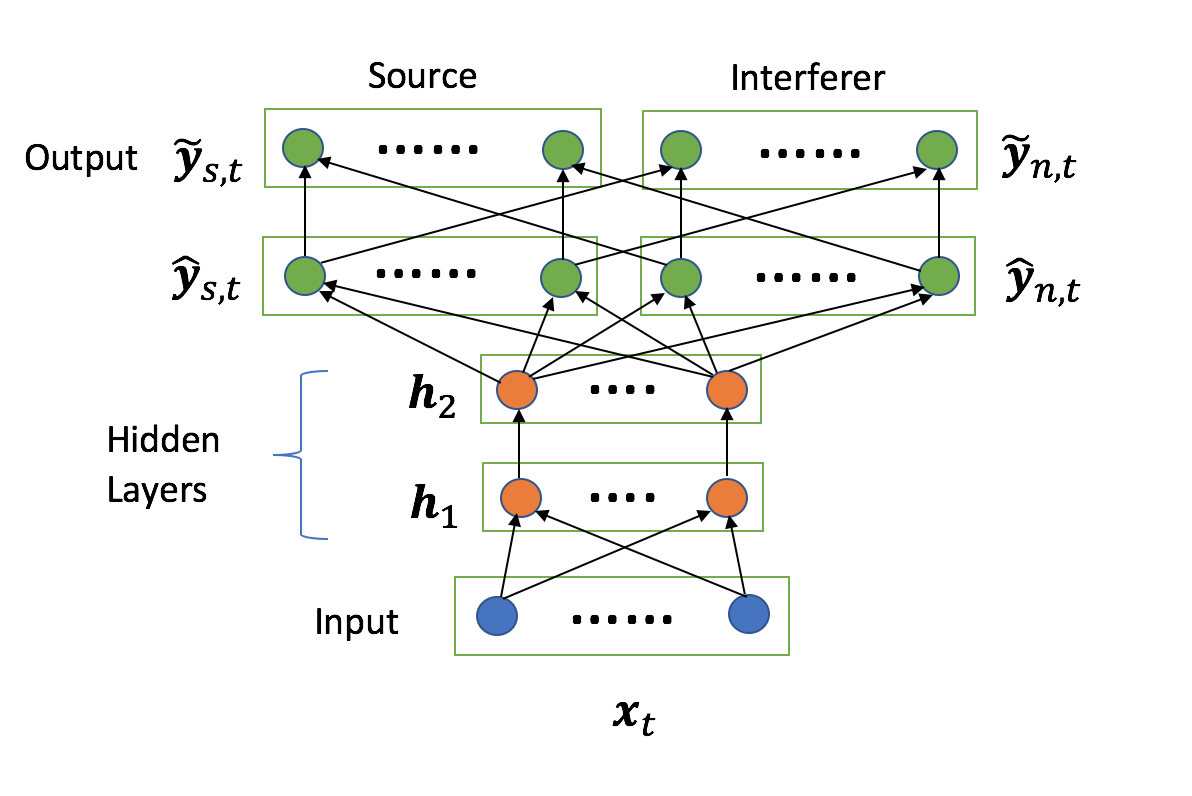}
	\caption{Separation using DF-DNN}
	\label{fig:nnet}
\end{figure}
\section{Our Approach}
\label{sec:algo}
We now discuss how to effectively solve a two-source separation problem using the proposed DF-DNN technique. 
\subsection{Joint Masking \cite{joint_mask}}
The features used to train the networks in this work are the magnitude of the Short Time Fourier Transform (STFT). Each signal is divided into $T$ frames and a $N$ point FFT is taken for each frame resulting in a $N\times T$ STFT matrix representing the signal. If the input to the network is the magnitude spectra of the mixture represented by $\boldsymbol{x}_t$ for time frame $t$ and the output predictions for a two source case are denoted by $\hat{\boldsymbol{y}}_{1,t}$ and $\hat{\boldsymbol{y}}_{2,t}$, then the masks are denoted by
\begin{equation}
\hat{\boldsymbol{m}}_{i,t} = \frac{|\hat{\boldsymbol{y}}_{i,t}|}{|\hat{\boldsymbol{y}}_{1,t}|+|\hat{\boldsymbol{y}}_{2,t}|}.
\end{equation}
Here, the multiplications and divisions are done element-wise. The method used in \cite{joint_mask} optimizes the masks along with the deep neural network parameters; see Fig. \ref{fig:nnet}. For this purpose, another layer is added at the output which corresponds to the masked predictions given by
\begin{equation}
\tilde{\boldsymbol{y}}_{i,t} = \frac{|\hat{\boldsymbol{y}}_{i,t}|}{|\hat{\boldsymbol{y}}_{1,t}|+|\hat{\boldsymbol{y}}_{2,t}|}\odot\boldsymbol{x}_t.
\end{equation}
This output layer is only dependent on the outputs of the previous layer $\hat{\boldsymbol{y}}_{i,t}$ and no weights are required to optimize for it. Given the output predictions $\tilde{\boldsymbol{y}}_1$ and $\tilde{\boldsymbol{y}}_2$, the network is trained to minimize the error between the predicted sources from the original sources. Also, for an effective separation, each predicted source should have maximum distance from the other source, that is, the network should be discriminative. To ensure the above conditions, the network is trained to optimize the following objective function
\begin{equation}\label{joint}
J = \frac{1}{2}(\|\boldsymbol{y}_1 - \tilde{\boldsymbol{y}}_1\|_F^2 + \|\boldsymbol{y}_2 - \tilde{\boldsymbol{y}}_2\|_F^2 - \gamma\|\boldsymbol{y}_1 - \tilde{\boldsymbol{y}}_2\|_F^2 - \gamma\|\boldsymbol{y}_2 - \tilde{\boldsymbol{y}}_1\|_F^2),
\end{equation}
where $\gamma$ controls the amount of discrimination the network can provide between the two sources. The predicted magnitude spectra $\tilde{\boldsymbol{y}}_1, \tilde{\boldsymbol{y}}_2$ and the phase of the mixture spectrum are used to create the STFT features of the separated sources. An inverse STFT operation results gives the corresponding time domain signals estimates $\hat{y}_i,~i=1,2$.
\subsection{Our Framework (DF-DNN)}
We denote the source signal by $\boldsymbol{y}_s$ and the (combined) interferer by $\boldsymbol{y}_n$. We are not concerned about the interference that can be introduced in estimated interferer from the source. Another hyper parameter $\mu$ is used here that controls the relative reconstruction of the source and the interferer. Further, since we are only concerned with the orthogonal component of the interferer not affecting the source, the modified formulation used in our framework is
\begin{equation}\label{df-dnn}
J = \frac{1}{2}(\|\boldsymbol{y}_s - \tilde{\boldsymbol{y}}_s\|_F^2 + \mu \|\boldsymbol{y}_n - \tilde{\boldsymbol{y}}_n\|_F^2 - \gamma\|\tilde{\boldsymbol{y}}_s - \boldsymbol{y}_{n,o}\|_F^2).
\end{equation}
Here $\boldsymbol{y}_{n,o}$ denotes the component of $\boldsymbol{y}_{n}$ orthogonal to the source subspace. Since the exact subspace of source is unknown to us, we can at best approximate it using the source training data $\boldsymbol{y}_{s}$. Hence, $\boldsymbol{y}_{n,o}$ is also an approximation of the orthogonal component of the interferer. Algorithm \ref{findorth} elaborates the computation of $\boldsymbol{y}_{n,o}$.
\begin{algorithm}[h]
	\caption{find\_orth}
	\label{findorth}
	\begin{algorithmic}[1]
		\STATE \textbf {Input}: $\boldsymbol{y}_{s}$, $\boldsymbol{y}_{n}$
		\STATE $\boldsymbol{y}_{s} = \boldsymbol{y}_{s} - \rm{mean}(\boldsymbol{y}_{s})$;
		\STATE $\boldsymbol{U},\boldsymbol{\Sigma},\boldsymbol{V} = \rm{SVD}(\boldsymbol{y}_{s})$;
		\STATE $\rm{total}\_\rm{energy} = \|\Sigma\|_F^2$
		\STATE $d =$ no. of columns containing $0.95*\rm{total}\_\rm{energy}$
		\STATE $\boldsymbol{S} = \boldsymbol{U}(:,1:d)$
		\STATE $\boldsymbol{y}_{n,o} = \boldsymbol{y}_{n} - P_{\boldsymbol{S}}\boldsymbol{y}_{n}$
		\STATE \textbf{Output}: $\boldsymbol{y}_{n,o}$
	\end{algorithmic}
\end{algorithm}
\subsection{Automatic Parameter Tuning}
\begin{figure}
	 \centering
	\includegraphics[height=3cm,width=\linewidth]{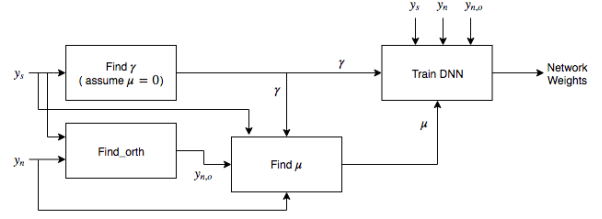}
	\caption{Block diagram for training DF-DNN}
	\label{fig:block}
\end{figure}

\begin{figure}
	\centering
	\begin{subfigure}{.45\linewidth}
		\centering
		\includegraphics[width=\linewidth]{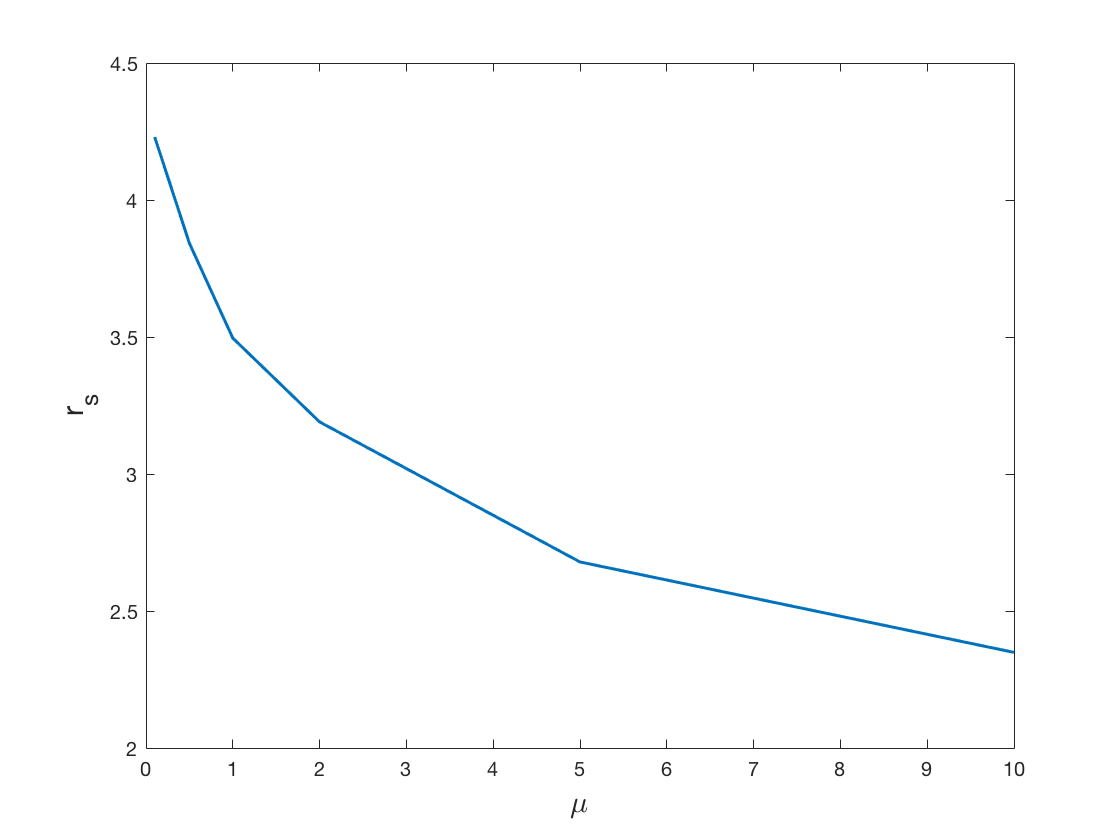}
		% 	\caption{Change of $r_s$ with $\mu$}
		% 	\label{fig:rs}
	\end{subfigure}
	\begin{subfigure}{.45\linewidth}
		\centering
		\includegraphics[width=\linewidth]{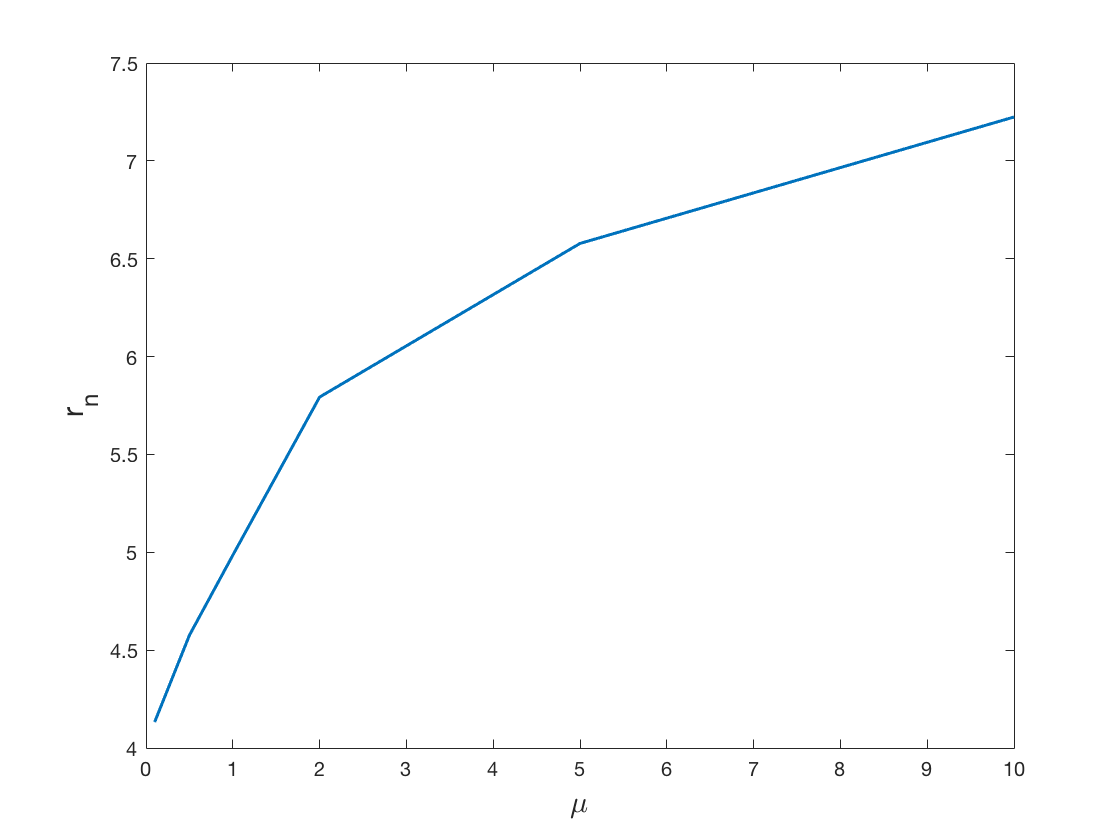}
		% 	\caption{Change of $r_n$ with $\mu$}
		% 	\label{fig:rn}
	\end{subfigure}
	\captionsetup{justification=centering}
	\caption{Change in ratios with $\mu$}
	\label{fig:ratios}
\end{figure}

Another modification that we propose is that the hyper-parameters $\gamma$ and $\mu$, which are fixed in joint separation, are automatically searched in our formulation \eqref{df-dnn} using the ratios $r_e, r_s, r_n$, we first defined in \cite{dfnmf}. The error ratio is defined as:
\begin{equation}\label{re}
r_e = \frac{\|\boldsymbol{y}_n - \tilde{\boldsymbol{y}}_{ns}\|_F}{\|\boldsymbol{y}_s - \tilde{\boldsymbol{y}}_{ss}\|_F},
\end{equation}
where $\tilde{\boldsymbol{y}}_{ns}$ and $\tilde{\boldsymbol{y}}_{ss}$ are the estimates of $\boldsymbol{y}_{s}$ when the input to the network is $\boldsymbol{y}_{n}$ and $\boldsymbol{y}_{s}$ respectively. $r_e$ is a measure of the interference that can be introduced in the source and a higher value of $r_e$ implies lesser interference. Hence, that value of $\gamma$ that gives the maximum $r_e$ is taken. The full procedure is described in Algorithm \ref{algo}. The energy ratios $r_s$ and $r_n$ are the ratios between energies in the source and interferer predictions when only only $\boldsymbol{y}_{s}$ and $\boldsymbol{y}_{n}$ respectively are given as inputs to the network, that is
\begin{align} \label{energy_ratios}
r_s = \frac{\|\tilde{\boldsymbol{y}}_{ss}\|_F}{\|\tilde{\boldsymbol{y}}_{sn}\|_F} &&
r_n = \frac{\|\tilde{\boldsymbol{y}}_{nn}\|_F}{\|\tilde{\boldsymbol{y}}_{ns}\|_F}.
\end{align}
Algorithm \ref{findmu} describes the usage of $r_s$ and $r_n$ to find the hyper-parameter $\mu$. For a fixed value of $\gamma$, the network is trained for successively increasing values of $\mu$. It is observed experimentally that the energy ratio $r_s$ monotonically decreases and $r_n$ increases with increasing values of $\mu$ as depicted in Figure \ref{fig:ratios}. The search is continued until $r_s$ is greater than $r_n$ and it remains more than a certain pre-decided threshold. The complete algorithm is described in Algorithm \ref{algo}. The steps involved in training of the auto-tuned network are shown in Figure \ref{fig:block}. Once the search for hyper-parameters is complete, the network is trained so that it minimizes the objective function \eqref{df-dnn}. The magnitude spectra of the test mixture $\boldsymbol{x}_{test}$ is then given as input to the trained network to get the predicted source output.
\begin{algorithm}[h]
	\caption{find\_mu}
	\label{findmu}
	\begin{algorithmic}[1]
		\STATE \textbf {Input}: $\boldsymbol{y}_{s}, \boldsymbol{y}_{n}, \boldsymbol{y}_{n,o}$, $\gamma$
		\STATE $\mu_{\rm{set}}$ = {set of possible values};
		\STATE $\rm{flag} = 0; k = 1;$
		\STATE $l = \rm{length}(\mu_{set})$
		\WHILE{flag == 0}
		\STATE $\mu = \mu_{\rm{set}}(k)$
		\STATE Train network with objective \eqref{df-dnn} with $\boldsymbol{x}$ as input features.
		\STATE Find $r_s,r_n$ as in \eqref{energy_ratios}
		\IF{$(L-1)r_s \leq r_{n} ~\algorithmicor~ r_s \leq r_{s,\rm{min}} ~\algorithmicor~ k==l$}
		\STATE flag = 1
		\ENDIF
		\STATE $k = k+1$
		\ENDWHILE
		\STATE \textbf {Output}: $\mu$
	\end{algorithmic}
\end{algorithm}
\begin{algorithm}[h]
	\caption{DF-DNN}
	\label{algo}
	\begin{algorithmic}[1]
		\STATE \textbf {Input}: $\boldsymbol{y}_{s}$, $\boldsymbol{y}_{n}$, $\boldsymbol{y}_{n,o}$, $\boldsymbol{x}_{test}$
		\STATE $\mu$ = 0;
		\STATE $\gamma$ = $\gamma_{\rm{min}}$;
		\WHILE{$\gamma_{\rm{min}} \leq \gamma \leq \gamma_{\rm{max}}$}
		\STATE Train the network with $\boldsymbol{y}_s$ as input features.
		\STATE Find $r_e$ as in \eqref{re}
		\STATE $\gamma = \gamma + 0.1$
		\ENDWHILE
		\STATE $\gamma$ = $\gamma$ corresponding to max($r_e$)
		\STATE $\mu$ = $\rm{find}\_\rm{mu}$($\boldsymbol{y}_s, \boldsymbol{y}_{n}, \boldsymbol{y}_{n,o}, \gamma$)
		\STATE Train a network with objective \eqref{df-dnn} 
		\STATE Input $\boldsymbol{x}_{test}$ into the network to get estimated source output $\tilde{\boldsymbol{y}}_{s}$
	\end{algorithmic}
\end{algorithm}

\section{Results}
\subsection{Data}
The proposed source separation strategy was tested to separate mixtures of 2, 3, and 4 speech signals. Two different datasets namely TIMIT \cite{timit} and TSP \cite{tsp} were used to create the mixtures, the signal to signal being 0 dB in each case. 8 male and 8 female speakers were taken from the TIMIT 16k dataset, which consists of 10 sentences per speaker. Nine sentences, which adds up to around 20 secs, were used up for training the networks and the remaining one was used as test case. The TSP dataset, sampled at 44.1k, consists of 60 sentences per speaker out of which 54 (about 125 secs) were used for training and remaining 6 were used as test cases. A total of 3 female and 3 male speakers were used from TSP dataset. 
\subsection{Parameters}
For the TIMIT dataset, framing of the signals was done using a Hamming window of length 512 with 50\% overlap and a 512 point FFT was taken for each frame. A two hidden layer network with 150 nodes per layer is used for this data. For the TSP dataset, 1024 length frames with 50\% overlap and then a 1024 point FFT for each frame was used to create the required features. The two layer networks used for this dataset had 300 nodes in each layer. For training of the networks, a batch size of 10000 frames were used. Each hidden unit used RELU as the activation function. 
\par
Algorithm \ref{findmu} uses a threshold $r_{s,\rm{min}}$ for the ratio based parameters to search for appropriate $\mu$. This threshold was set to be 8 for both the datasets. The $\mu_{set}$ used in Algorithm \ref{findmu} is $\{0.1,0.5,1,2,5,10\}$. The limits for $\gamma$ in Algorithm \ref{algo} are taken to be $\gamma_{\rm{min}} = 0.1$ and $\gamma_{\rm{max}} = 0.5$.
\subsection{Performance}
The performance of the proposed framework is evaluated on the basis of SDR, SIR and SAR metric calculated using the BSS evaluation toolbox \cite{perf_eval}. Our experiments show that training a network using DF-DNN, which separates one component at a time out of the mixture while targeting the orthogonal component of the combined interferer using \eqref{df-dnn} as the objective function and performing a search for the suitable hyper-parameters, results in better SDR and SAR while keeping the SIR nearly similar as compared to training the network for joint separation of all the sources underlying the mixture using a single network trained with \eqref{joint} as the objective function. SAR indicates the amount of artifacts introduced in the separated source. Since DF-DNN framework focuses on proper separation of only one source at a time, it is evidently better than joint separation in terms of reconstruction. Also as apparent from the numbers in the tables \ref{table:timit} and \ref{table:tsp}, SIR is nearly same or slightly lower for all separation cases. This is attributed to the imperfection in computing the orthogonal component of interferer $\boldsymbol{y}_{n,o}$ as a result of which some interference may be introduced in the recovered source. The overall distortion, accounted for in SDR, is however improved.
\begin{table}[h!]
	\caption{\it Average performance for TIMIT dataset}
	\label{table:timit}
	\centering
	{
		\begin{tabular}{|c|c| c| c|}
			\hline \Tstrut
			&&	DF-DNN & Joint\\
			\hline \Tstrut
			\multirow{3}{*}{2 sources}%
			&SDR & 6.39 & 5.36  \\ 
			&SIR & 9.72 & 9.226 \\
			&SAR & 9.89 & 8.57\\
			\hline \Tstrut
			\multirow{3}{*}{3 sources}%
			&SDR & 2.62 & 2.29\\
			&SIR & 5.77 & 5.87 \\
			&SAR & 6.97 & 6.1\\
			\hline \Tstrut
			\multirow{3}{*}{4 sources}%
			&SDR & 0.07 & -1.107 \\
			&SIR & 3.19 & 2.54 \\
			&SAR & 5.08 & 3.84 \\
			\hline
		\end{tabular}
	}
\end{table}
\begin{table}[h!]
	\caption{\it Average performance for TSP dataset}
	\label{table:tsp}
	\centering {
		\begin{tabular}{|c|c| c| c|}
			\hline \Tstrut
			&&	DF-DNN & Joint\\
			\hline \Tstrut
			\multirow{3}{*}{2 sources}%
			&SDR & 7.19 & 6.65  \\ 
			&SIR & 12.02 & 12.26 \\
			&SAR & 9.34 & 8.44\\
			\hline \Tstrut
			\multirow{3}{*}{3 sources}%
			&SDR & 2.19 & 1.06\\
			&SIR & 6.4 & 6.61 \\
			&SAR & 5.47 & 3.58\\
			\hline \Tstrut
			\multirow{3}{*}{4 sources}%
			&SDR & 0.42 & -0.422 \\
			&SIR & 4.57 & 5.22 \\
			&SAR & 4.14 & 2.4 \\
			\hline
		\end{tabular}
	}
\end{table}

\section{Conclusion}
We present a discriminative neural network based framework for the SCSS problem
where we target an automated training system that can tune its parameters according to the mixture it wants to separate. The framework also focuses on only one source at a time thereby improving its reconstruction. We applied our framework on a two layer network
and through experiments show that our method improves the separation performance compared to joint separation strategy. 
\bibliographystyle{IEEEtran}

\bibliography{mybib}

% \begin{thebibliography}{9}
% \bibitem[1]{Davis80-COP}
%   S.\ B.\ Davis and P.\ Mermelstein,
%   ``Comparison of parametric representation for monosyllabic word recognition in continuously spoken sentences,''
%   \textit{IEEE Transactions on Acoustics, Speech and Signal Processing}, vol.~28, no.~4, pp.~357--366, 1980.
% \bibitem[2]{Rabiner89-ATO}
%   L.\ R.\ Rabiner,
%   ``A tutorial on hidden Markov models and selected applications in speech recognition,''
%   \textit{Proceedings of the IEEE}, vol.~77, no.~2, pp.~257-286, 1989.
% \bibitem[3]{Hastie09-TEO}
%   T.\ Hastie, R.\ Tibshirani, and J.\ Friedman,
%   \textit{The Elements of Statistical Learning -- Data Mining, Inference, and Prediction}.
%   New York: Springer, 2009.
% \bibitem[4]{YourName17-XXX}
%   F.\ Lastname1, F.\ Lastname2, and F.\ Lastname3,
%   ``Title of your INTERSPEECH 2018 publication,''
%   in \textit{Interspeech 2018 -- 19\textsuperscript{th} Annual Conference of the International Speech Communication Association, September 2-6, Hyderabad, India Proceedings, Proceedings}, 2018, pp.~100--104.
% \end{thebibliography}

\end{document}